\documentstyle[preprint,aps,epsf]{revtex}
\def\simgt{\,\rlap{\lower 3.5 pt\hbox{$\mathchar \sim$}}\raise 1pt \hbox {$>$}\,}
\def\simlt{\,\rlap{\lower 3.5 pt\hbox{$\mathchar \sim$}}\raise 1pt \hbox {$<$}\,}
\begin{document}

\draft
\tightenlines

\title{
\vspace*{-35pt}
{\normalsize \hfill {\sf BI-TP 2003/38}} \\
Remarks on the multi-parameter reweighting method for 
the study of lattice QCD at non-zero temperature and density
}
\author{Shinji Ejiri}
\address{Fakult\"at f\"ur Physik, Universit\"at Bielefeld,
          D-33615 Bielefeld, Germany} 

\date{\today}
\maketitle

\begin{abstract}
We comment on the reweighting method for the study of finite density 
lattice QCD. 
We discuss the applicable parameter range of the reweighting method for 
models which have more than one simulation parameter. 
The applicability range is determined by the fluctuations of the modification 
factor of the Boltzmann weight.
In some models having a first order phase transition, 
the fluctuations are minimized along the phase transition line 
if we assume that the pressure in the hot and the cold phase is balanced 
at the first order phase transition point.
This suggests that the reweighting method with two parameters is 
applicable in a wide range for the purpose of tracing out the phase 
transition line in the parameter space. 
To confirm the usefulness of the reweighting method for 2 flavor QCD, 
the fluctuations of the reweighting factor are measured by numerical 
simulations for the cases of reweighting in the quark mass and 
chemical potential directions. The relation with the phase 
transition line is discussed. Moreover, the sign problem 
caused by the complex phase fluctuations is studied.
\end{abstract}




\section{Introduction}
\label{sec:intro}

The study of the phase structure of QCD at non-zero temperature $T$ 
and non-zero quark chemical potential $\mu_q$ is currently one of the most 
attractive topics in particle physics \cite{KL03,MNNT}. 
The heavy-ion collision experiments aiming to produce the 
quark-gluon plasma (QGP) are running at BNL and CERN, 
for which the interesting regime is rather low density.  
Moreover a new color superconductor phase is expected 
in the region of low temperature and high density. 
In the last few years, remarkable progress has been achieved in 
the numerical study by Monte-Carlo simulations of lattice QCD 
at low density.
It was shown that the phase transition line, separating hadron phase 
and QGP phase, can be traced out from $\mu_q=0$ to finite $\mu_q$, 
and it was also possible to investigate the equation of state 
quantitatively at low density. 
The main difficulty of the study at non-zero baryon density is that 
the Monte-Carlo method is not applicable directly at finite density, 
since the fermion determinant is complex for non-zero $\mu_q$ and 
configurations cannot be generated with the probability 
of the Boltzmann weight. The most popular technique to study at 
non-zero $\mu_q$ is the reweighting method; performing simulations at 
${\rm Re}(\mu_q)=0$, and then modify the Boltzmann weight at the step of 
measurement of observables \cite{FK,us02,Christian,FKS}. 
The Glasgow method \cite{Glasgow} is one of the reweighting methods. 
A composite (Glasgow) reweighting method has recently been proposed 
by \cite{PC}. 
Another approach is analytic continuation from simulations 
at imaginary chemical potential \cite{Roberge,dFP,dEL}. 
Moreover, calculating coefficients of a Taylor expansion in terms of $\mu_q$ 
is also a hopeful approach for the study at non-zero baryon density 
\cite{us02,QCDTARO,us03,GG}. 
The studies by Taylor expansion or imaginary chemical potential require 
analyticity of physical quantities as functions of $T$ and $\mu_q$, 
while the reweighting method has a famous ``sign problem". 
The sign problem is caused by complex phase fluctuations of the fermion 
determinant, which are measured explicitly in Ref.~\cite{dFKT}, and 
also Ref.~\cite{AANV} is a trial to avoid the sign problem. 

In this paper, 
we make comments on the reweighting method with respect to more than one 
simulation parameter, particularly, including a chemical potential. 
Because the fluctuation of the modification factor (reweighting factor) 
enlarges the statistical error, the applicable range of this method is 
determined by the fluctuation of the reweighting factor. 
When the error becomes considerable in comparison with the expectation 
value, the reweighting method breaks down. 
An interesting possibility is given in the following case: 
when we change two parameters at the same time, the reweighting factors 
for two parameters might cancel each other, 
then the error does not increase and also the expectation values of 
physical quantities do not change by this parameter shift. Therefore 
finding such a direction provides useful information for mapping out the value 
of physical quantities in the parameter space.
As we will see below, there is such a direction in the parameter space and 
the knowledge of this property of the reweighting method makes the method 
more useful. 
Fodor and Katz \cite{FK} investigated the phase transition line for rather 
large $\mu_q$. This argument may explain why they could 
calculate $\beta_c$ for such large $\mu_q$.

In the next section, we explain the reweighting method with multi-parameters. 
Then, in Sec.~\ref{sec:anisoquench}, the case of SU(3) pure gauge theory 
on an anisotropic lattice is considered as the simplest example, 
and we proceed to full QCD with a first order phase transition 
in Sec.~\ref{sec:full1st}. 
For 2-flavor QCD, the reweighting method with respect to quark mass is 
discussed in Sec.~\ref{sec:massdir}. The application to non-zero baryon 
density is discussed in Sec.~\ref{sec:mudir}. The problem of the complex measure 
is also considered in Sec.~\ref{sec:phase}. 
Conclusions are given in Sec.~\ref{sec:concl}.

\section{Reweighting method and the applicability range}
\label{sec:method}

The reweighting method is based on the following identity:
\begin{eqnarray}
\langle {\cal O} \rangle_{(\beta, m, \mu)}
&=& \frac{1}{{\cal Z}_{(\beta, m, \mu)}} \int 
D U {\cal O} ( \det M(m, \mu))^{N_{\rm f}} e^{-S_g(\beta)} 
\nonumber
\\
&=& \frac{\left\langle {\cal O} e^{F} e^{G}
 \right\rangle_{(\beta_0, m_0, \mu_0)} }{\left\langle e^{F} e^{G}
  \right\rangle_{(\beta_0, m_0, \mu_0)}} 
= \frac{\left\langle {\cal O} e^{\Delta F} e^{\Delta G}
 \right\rangle_{(\beta_0, m_0, \mu_0)} }{\left\langle e^{\Delta F} 
e^{\Delta G}  \right\rangle_{(\beta_0, m_0, \mu_0)}}. 
\label{eq:rew} 
\end{eqnarray}
Here $M$ is the quark matrix, $S_g$ is the gauge action, 
$N_{\rm f}$ is the number of flavors 
($N_{\rm f}/4$ for staggered type fermions instead of $N_{\rm f}$), 
$F=N_{\rm f}[\ln \det M(m, \mu)-\ln \det M(m_0, \mu_0)],$
$G=(\beta-\beta_0) P, P=-\partial S_g/\partial \beta,$ 
$\beta=6/g^2,$ and 
$\Delta F[G] = F[G] - \langle F[G] \rangle$.
$m$ and $\mu \equiv \mu_q a$ are the bare quark mass and 
chemical potential in a lattice unit, respectively.
The expectation value $\langle {\cal O} \rangle_{(\beta, m, \mu)}$ 
can, in principle, be computed by simulation at $(\beta_0, m_0, \mu_0)$ 
using this identity \cite{Swen88}.
However, a problem of the reweighting method is that the fluctuation of 
$e^{\Delta F} e^{\Delta G} =e^F e^G /(e^{\langle F \rangle} 
e^{\langle G \rangle})$ enlarges the statistical error of the numerator 
and denominator of Eq.(\ref{eq:rew}). The worst case, which is called 
``Sign problem", is that the sign of the reweighting factor changes frequently 
during Monte-Carlo steps, then the expectation values in 
Eq.(\ref{eq:rew}) become vanishingly small in comparison with the error, and 
this method does not work. 
However, the fluctuations cause the $\beta$, $m$ and $\mu$ dependence of 
$\langle {\cal O} \rangle_{(\beta, m, \mu)}$. 
Otherwise, if $ e^F e^G$ does not fluctuate, 
$ e^{\Delta F} e^{\Delta G}=1$ and 
$\langle {\cal O} \rangle_{(\beta, m, \mu)}$ does not change with 
parameter change. 
Roughly speaking, the difference of 
$\langle {\cal O} \rangle_{(\beta, m, \mu)}$ 
from $\langle {\cal O} \rangle_{(\beta_0, m_0, \mu_0)}$ increases as 
the magnitude of fluctuations of $F$ and $G$ increases and, 
if $F$ and $G$ have a correlation, the increase of the total 
fluctuation as a function of $\beta$, $m$ and $\mu$ is non-trivial. 
Therefore, it is important to discuss the correlation between $e^{F}$ and 
$ e^{G}$ and to estimate the total fluctuation of the reweighting factor 
in the parameter space in order to estimate the applicability range 
of the reweighting method in the parameter space and 
how the system changes by parameter shifts. 
This is helpful information for the study of QCD thermodynamics.

\section{SU(3) gauge theory on an anisotropic lattice}
\label{sec:anisoquench}

Let us start with the case of 
SU(3) pure gauge theory on an anisotropic lattice, 
having two different lattice spacings for space and time directions: 
$a_{\sigma}$ and $a_{\tau}$. 
As we will show for this case, there is a clear relation 
between the phase transition line and the direction which 
minimizes the fluctuation of the reweighting factor. 
The action is 
\begin{eqnarray}
 S_g = -\beta_{\sigma} \sum_x P_{\sigma}(x)-\beta_{\tau} \sum_x P_{\tau}(x),
\end{eqnarray}
where $ P_{\sigma (\tau)}$ is spatial (temporal) plaquette. 
The SU(3) pure gauge theory has a first order phase transition. 
At the transition point $(T_c)$, there exist two phases simultaneously. 
For the two phases to coexist, the pressure in these phases must be equal: 
$\Delta p \equiv p^{\rm (hot)} - p^{\rm (cold)} = 0$. 
If we require $\Delta p =0$, we find that the phase transition line 
in the parameter space of $(\beta_{\sigma}, \beta_{\tau})$ has to run 
in such a direction that 
the fluctuation of the reweighting factor is minimized when 
we perform a simulation on the phase transition point and apply the 
reweighting in the $(\beta_{\sigma}, \beta_{\tau})$ plane. 

A significant feature of a Monte-Carlo simulation at a first order phase 
transition point is the occurrence of flip-flops between configurations of hot and cold phases.
If one write a histogram of the action density, i.e. the plaquette: 
$P_{\sigma}$ and $P_{\tau}$, there exist two peaks. 
The value of the action density changes sometimes from one near one peak 
to one near the other peak during Monte-Carlo steps \cite{FOU,QCDPAX}. 
The flip-flop is the most important fluctuation; 
in fact, the flip-flop makes the strong peak of susceptibilities of 
observables such as the plaquette or the Polyakov loop at the transition point.
Also, the flip-flop implies strong correlations between $P_{\sigma}$ and 
$ P_{\tau}$ because the values of $P_{\sigma}$ and $P_{\tau}$ change 
simultaneously between the typical values of the two phases in the 
$(P_{\sigma}, P_{\tau})$ plane. 

Here, we discuss the fluctuation of the reweighting factor when one 
performs a simulation at the transition point,
$(\beta_{\sigma 0}, \beta_{\tau 0})$. 
An expectation value of ${\cal O}$ at $(\beta_{\sigma}, \beta_{\tau})$ 
on an anisotropic lattice is calculated by 
\begin{eqnarray}
\langle {\cal O} \rangle_{(\beta_{\sigma}, \beta_{\tau})}
= \left. \left\langle {\cal O} e^{-\Delta S_g}
 \right\rangle_{(\beta_{\sigma 0}, \beta_{\tau 0})} \right/ 
\left\langle e^{-\Delta S_g} 
\right\rangle_{(\beta_{\sigma 0}, \beta_{\tau 0})}, 
\end{eqnarray}
where $\Delta S_g = -\Delta \beta_{\sigma} \sum_x P_{\sigma}(x) 
-\Delta \beta_{\tau} \sum_x P_{\tau}(x),$ and $\Delta \beta_{\sigma (\tau)} 
= \beta_{\sigma (\tau)} -\beta_{\sigma (\tau) 0}.$ 
For simplification, we ignore the local fluctuation around the two peaks of 
the histogram of the action density and consider only the flip-flop 
between hot and cold phases, since it is the most important fluctuation 
at the first order phase transition point. 
Then, the fluctuation is estimated 
by the difference of the reweighting factor between hot and cold phases,
up to first order, 
\begin{eqnarray}
|e^{-\Delta S_g^{\rm (hot)}}-e^{-\Delta S_g^{\rm (cold)}}| \approx
3 N_{\rm site}| \Delta \beta_{\sigma} 
(\bar{P}_{\sigma}^{\rm (hot)}-\bar{P}_{\sigma}^{\rm (cold)})
+ \Delta \beta_{\tau} 
(\bar{P}_{\tau}^{\rm (hot)}-\bar{P}_{\tau}^{\rm (cold)}) | +\cdots,
\label{eq:sgflc}
\end{eqnarray}
where $\bar{P}_{\sigma}^{\rm hot (cold)}$ and 
$\bar{P}_{\tau}^{\rm hot (cold)}$ are 
the average values of the spatial and temporal plaquettes for 
configurations in the hot (cold) phase and 
$N_{\rm site} \equiv N_{\sigma}^3 \times N_{\tau}$ 
is the number of sites for an $N_{\sigma}^3 \times N_{\tau}$ lattice. 
Hence, along the line which has a slope 
\begin{eqnarray}
\frac{{\rm d} \beta_{\sigma}}{{\rm d} \beta_{\tau}} = 
- \frac{\bar{P}_{\tau}^{\rm (hot)}-\bar{P}_{\tau}^{\rm (cold)}}
{\bar{P}_{\sigma}^{\rm (hot)}-\bar{P}_{\sigma}^{\rm (cold)}}, 
\label{eq:dirmin}
\end{eqnarray}
the fluctuation of the reweighting factor is canceled to leading order. 

On the other hand, since $V=(N_{\sigma} a_{\sigma})^3$ and 
$T=(N_{\tau} a_{\tau})^{-1}$, pressure is defined by 
\begin{eqnarray}
p &=& T \left. \frac{\partial \ln {\cal Z}}{\partial V} \right|_T
= \frac{1}{3N_{\sigma}^3 N_{\tau} a_{\sigma}^2 a_{\tau}} \left. 
\frac{\partial \ln {\cal Z}}{\partial a_{\sigma}} \right|_{a_{\tau}}, \\
\frac{p}{T^4} &=& 
N_{\tau}^4 \left( \frac{a_{\tau}}{a_{\sigma}} \right)^3 \left[ 
a_{\sigma} \frac{\partial \beta_{\sigma}}{\partial a_{\sigma}} 
(\langle \bar{P}_{\sigma} \rangle - \langle \bar{P}_{\sigma} \rangle_{0})  
+ a_{\sigma} \frac{\partial \beta_{\tau}}{\partial a_{\sigma}} 
(\langle \bar{P}_{\tau} \rangle - \langle \bar{P}_{\tau} \rangle_{0}) \right],
\label{eq:press}
\end{eqnarray}
where 
$\bar{P}_{\sigma(\tau)}=(3N_{\rm site})^{-1}\sum_x P_{\sigma(\tau)}(x)$, 
and $\langle \bar{P}_{\sigma(\tau)} \rangle_{0}$ 
is the expectation value of the plaquette on a $T=0$ lattice 
for the normarization. 

By separating the configurations into those in hot and cold phases 
\cite{FOU,QCDPAX},
the gap of pressure between hot and cold phases at $T_c$ is computed by 
\begin{eqnarray}
\Delta \frac{p}{T^4} & \equiv & 
\frac{p^{\rm (hot)}}{T^4} -\frac{p^{\rm (cold)}}{T^4} \nonumber \\
&=& N_{\tau}^4 \left( \frac{a_{\tau}}{a_{\sigma}} \right)^3 \left[ 
a_{\sigma} \frac{\partial \beta_{\sigma}}{\partial a_{\sigma}} 
(\bar{P}_{\sigma}^{\rm (hot)} 
- \bar{P}_{\sigma}^{\rm (cold)})  
+ a_{\sigma} \frac{\partial \beta_{\tau}}{\partial a_{\sigma}} 
(\bar{P}_{\tau}^{\rm (hot)} 
- \bar{P}_{\tau}^{\rm (cold)}) \right]. 
\label{eq:pgap}
\end{eqnarray}
Since the gap of pressure should vanish, $\Delta p =0$, 
\begin{eqnarray}
\left. \frac{\partial \beta_{\sigma}}{\partial a_{\sigma}} \right/ 
\frac{\partial \beta_{\tau}}{\partial a_{\sigma}} 
= -\frac{\bar{P}_{\tau}^{\rm (hot)}-\bar{P}_{\tau}^{\rm (cold)}}
{\bar{P}_{\sigma}^{\rm (hot)}-\bar{P}_{\sigma}^{\rm (cold)}}. 
\end{eqnarray}
Moreover, because $T_c=(N_{\tau} a_{\tau})^{-1}$ on the phase 
transition line, $a_{\tau}$ keeps constant with $(N_{\tau} T_c)^{-1}$ 
along the transition line, i.e. 
\begin{eqnarray}
\Delta a_{\tau} = 
\frac{\partial a_{\tau}}{\partial \beta_{\sigma}} \Delta \beta_{\sigma} + 
\frac{\partial a_{\tau}}{\partial \beta_{\tau}} \Delta \beta_{\tau} =0, 
\end{eqnarray}
when one changes parameters 
$(\beta_{\sigma}, \beta_{\tau}) \to 
(\beta_{\sigma} +\Delta \beta_{\sigma}, \beta_{\tau} +\Delta \beta_{\tau})$ 
along the phase transition line.
Then the slope of the phase transition line $(r_t)$ \cite{Ej98} is obtained by 
\begin{eqnarray}
r_t \equiv 
\left. \frac{{\rm d} \beta_{\sigma}}{{\rm d} \beta_{\tau}} \right|_{T_c}
= \left. - \frac{\partial a_{\tau}}{\partial \beta_{\tau}} \right/ 
\frac{\partial a_{\tau}}{\partial \beta_{\sigma}} 
= \left. \frac{\partial \beta_{\sigma}}{\partial a_{\sigma}} \right/ 
\frac{\partial \beta_{\tau}}{\partial a_{\sigma}}, 
\end{eqnarray}
where we used an identify: 
\begin{eqnarray}
\left( \begin{array}{cc}
\frac{\partial \beta_{\sigma}}{\partial a_{\sigma}} &
\frac{\partial \beta_{\tau}}{\partial a_{\sigma}} \\
\frac{\partial \beta_{\sigma}}{\partial a_{\tau}} &
\frac{\partial \beta_{\tau}}{\partial a_{\tau}}
\end{array} \right) =
\frac{1}{
\frac{\partial a_{\tau}}{\partial \beta_{\tau}}
\frac{\partial a_{\sigma}}{\partial \beta_{\sigma}}
-\frac{\partial a_{\tau}}{\partial \beta_{\sigma}}
\frac{\partial a_{\sigma}}{\partial \beta_{\tau}}}
\left( \begin{array}{cc}
 \frac{\partial a_{\tau}}{\partial \beta_{\tau}} &
-\frac{\partial a_{\tau}}{\partial \beta_{\sigma}} \\
-\frac{\partial a_{\sigma}}{\partial \beta_{\tau}} &
 \frac{\partial a_{\sigma}}{\partial \beta_{\sigma}}
\end{array} \right).
\end{eqnarray}
Hence, the condition for $\Delta p=0$ becomes 
\begin{eqnarray}
\frac{\bar{P}_{\tau}^{\rm (hot)}-\bar{P}_{\tau}^{\rm (cold)}}
{\bar{P}_{\sigma}^{\rm (hot)}-\bar{P}_{\sigma}^{\rm (cold)}} = -r_t.
\label{eq:dp0cond}
\end{eqnarray}

This equation for ${\rm d}\beta_{\sigma}/{\rm d}\beta_{\tau}$ may 
correspond to the Clausius-Clapeyron equation in the $(p, T)$ plane: 
${\rm d}p/{\rm d}T=\Delta S/ \Delta V$ ($S$: entropy). 
In fact, Eq.(\ref{eq:dp0cond}) and the vanishing pressure gap, 
$\Delta p=0,$ are confirmed by calculating the slope of the transition 
line from the peak position of the Polyakov loop susceptibility 
obtained by numerical simulations in Ref.~\cite{Ej98}. 
Historically, the non-zero gap of pressure at the transition point 
had been a problem for long time. 
The reason of $\Delta p \neq 0$ was that the 
precise non-perturbative measurement of the anisotropy coefficients, 
$a_{\sigma} (\partial \beta_{\sigma}/\partial a_{\sigma}), 
a_{\sigma} (\partial \beta_{\tau}/\partial a_{\sigma}),$ etc., 
had been difficult, and the perturbative value \cite{Karsch} 
cannot be used at the phase transition point for $N_{\tau}=4$ or $6$. 
After the precise measurement of the anisotropy coefficients became 
possible, the problem of a non-zero pressure gap was solved. 
Also, determinations of these coefficients by another non-perturbative 
method have been done in Refs.~\cite{burgers,EKS,klassen}. 

\begin{figure}[t]
\centerline{
\epsfxsize=10.0cm\epsfbox{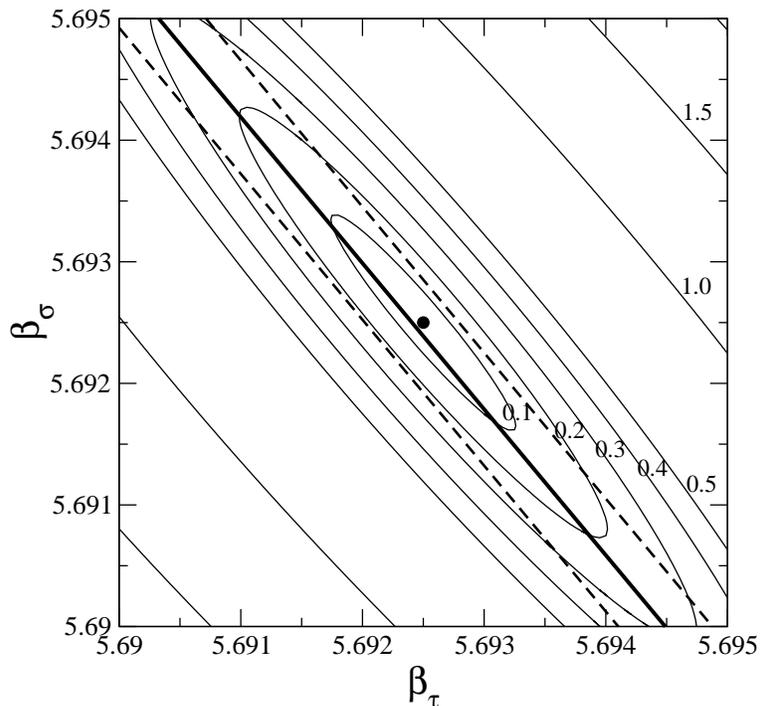}
}
\caption{
Contour plot of the standard deviation of the reweighting factor 
and the phase transition line in the $(\beta_{\sigma}, \beta_{\tau})$ plane. 
Bold line is the phase transition line, and the dashed lines denote its error. 
Values in this figure are the standard deviation divided by the mean value. 
The simulation point is $\beta=5.6925$.
}
\vspace*{-4mm}
\label{fig:conpure}
\end{figure}

From Eq.(\ref{eq:dirmin}) and Eq.(\ref{eq:dp0cond}), 
we find that the direction which minimizes the fluctuation of 
the reweighting factor must be the same as the direction of 
the phase transition line in the $(\beta_{\sigma}, \beta_{\tau})$ plane, 
if pressure in hot and cold phases are balanced at $T_c$, $\Delta p=0$. 
Here, in practice, we estimate the fluctuation of $e^{-\Delta S_g}$ by 
a numerical simulation. 
We compute the standard deviation of the reweighting factor using 
the data obtained in Ref.~\cite{QCDPAX}. 
The lattice size is $24^2 \times 36 \times 4$.
The data is generated by the standard Wilson gauge action with 
$\beta_{\sigma}=\beta_{\sigma}=5.6925$, 
that is just on the transition line, 
$\beta_c=5.69245(23)$ for $N_{\tau}=4$ at $a_{\sigma}=a_{\tau}$. 
Ellipses in Fig.\ref{fig:conpure} are the contour lines of 
the standard deviation normalized by the mean value: 
$\sqrt{ \langle (e^{-\Delta S_g})^2 \rangle 
- \langle e^{-\Delta S_g} \rangle^2}/\langle e^{-\Delta S_g} \rangle, $
and we write this value in Fig.~\ref{fig:conpure}.
We also denote the phase transition line, obtained by the measurement of 
the Polyakov loop susceptibility assuming that the peak position of the 
susceptibility is the phase transition point \cite{Ej98}, 
by a bold line, and dashed lines 
are the upper bound and lower bound. 
We find that the phase transition line and the line 
which minimizes fluctuations are consistent. 
This result also means that the reweighting method is applicable 
in a wide range 
for the determination of the phase transition line in the parameter space 
of SU(3) gauge theory on an anisotropic lattice, since 
the increase of the statistical error caused by the reweighting 
is small along the transition line.

We note that, from Eq.(\ref{eq:dp0cond}), $\Delta S_g^{\rm (hot)}$ and 
$\Delta S_g^{\rm (cold)}$ in Eq.(\ref{eq:sgflc}) are equal under the 
change along the phase transition line, hence the fluctuations are 
canceled in every order of $\Delta \beta_{\sigma (\tau)}$. 
For SU(3) pure gauge theory on an anisotropic lattice, 
the system is independent of $a_{\sigma}/a_{\tau}$ in a physical unit, 
hence the system does not change along the transition line except 
for the volume, $V=(N_{\sigma} a_{\sigma})^3$, 
if $N_{\sigma}$ is finite.\footnote{
In fact, as expected from the finite size scaling, 
the peak height of the Polyakov loop susceptibility increases 
as $a_{\sigma}$ increases \cite{Ej98}.}
Because physical quantities does not change without the fluctuation of 
reweighting factor, this result is quite natural.

\section{Full QCD with a first order phase transition}
\label{sec:full1st}

Next, we extend this discussion for the case of full QCD with a first 
order phase transition such as 3 flavor QCD near the chiral limit. 
The reweighting method is applied in the parameter space of $(m, \beta)$.
We consider Helmholtz free energy density $f =-T \ln {\cal Z} /V$ 
for a canonical ensemble which is equal to minus pressure, $p=-f$, for a 
large homogeneous system. 
Under a parameter change from $(m, \beta)$ to 
$(m+\Delta m, \beta +\Delta \beta)$, the variation of the free energy is
given, up to the first order, by 
\begin{eqnarray}
\left. \frac{f}{T^4}\right|_{(m+\Delta m, \beta +\Delta \beta)}- 
\left. \frac{f}{T^4}\right|_{(m, \beta)} 
= -N_{\tau}^4 \left[
(\langle \bar{Q}_1 \rangle - \langle \bar{Q}_1 \rangle_{0})  \Delta m 
+ 6(\langle \bar{P} \rangle - \langle \bar{P} \rangle_{0}) \Delta \beta 
\right] + \cdots
\label{eq:fe}
\end{eqnarray}
where 
$\bar{Q}_1=N_{\rm site}^{-1} N_{\rm f} \partial (\ln \det M)/ \partial m$ and 
$\bar{P}=-(6N_{\rm site})^{-1} \partial S_g/\partial \beta$. 
For the normalization at $T=0$, we subtract the zero temperature contribution, 
$\langle \bar{P} \rangle_{0}$ and $\langle \bar{Q}_1 \rangle_{0}$. 
Here, we should notice that the first derivatives of the free energy are 
discontinuous at the phase transition line, hence we cannot estimate 
the difference of the free energy beyond the transition line by this equation.

We assume that the gap of the pressure is zero in the entire parameter space 
$(m, \beta)$. 
We change $m$ and $\beta$ along the phase transition line starting at two 
points: 
just above and just below $\beta_c$, without crossing the transition line. 
Then the change of the free energy must be the same for both these cases, 
since a pressure gap is not generated under this variation, i.e. 
$\Delta(\Delta p)=0$. 
Hence, up to first order of $\Delta m$ and $\Delta \beta$, 
\begin{eqnarray}
&& \left( \left. 
\frac{f}{T^4}\right|_{(m+\Delta m, \beta+\Delta \beta)}^{(\rm hot)} 
-\left. \frac{f}{T^4}\right|_{(m, \beta)}^{(\rm hot)} 
\right) -\left( 
\left. \frac{f}{T^4}\right|_{(m+\Delta m, \beta+\Delta \beta)}^{(\rm cold)}  
-\left. \frac{f}{T^4}\right|_{(m, \beta)}^{(\rm cold)} 
\right) \nonumber \\
&& = -N_{\tau}^4 \left[
(\bar{Q}_1^{\rm (hot)} - \bar{Q}_1^{\rm (cold)}) \Delta m 
+ 6(\bar{P}^{\rm (hot)} - \bar{P}^{\rm (cold)}) \Delta \beta \right] 
+ \cdots =0.
\label{eq:diff}
\end{eqnarray}
is required on the first order phase transition line. 
From this equation, we obtain a similar relation as Eq.(\ref{eq:dp0cond}): 
\begin{eqnarray}
 \frac{\bar{Q}_1^{\rm (hot)} - \bar{Q}_1^{\rm (cold)}}
{6(\bar{P}^{\rm (hot)} - \bar{P}^{\rm (cold)})}
= -\left. \frac{{\rm d} \beta}{{\rm d} m} \right|_{T_c}. 
\label{eq:p0full}
\end{eqnarray}

On the other hand, the change of the reweighting factor under 
flip-flop is 
\begin{eqnarray}
&& |e^{\Delta F^{\rm (hot)}} e^{\Delta G^{\rm (hot)}} 
-e^{\Delta F^{\rm (cold)}} e^{\Delta G^{\rm (cold)}}| \nonumber \\
&& \approx N_{\rm site} \left[
(\bar{Q}_1^{\rm (hot)} - \bar{Q}_1^{\rm (cold)}) \Delta m 
+ 6(\bar{P}^{\rm (hot)} - \bar{P}^{\rm (cold)}) \Delta \beta \right] 
+ \cdots .
\label{eq:flipfull}
\end{eqnarray}
If we ignore the local fluctuation around the peaks of the 
distribution of $P$ and $Q_1$, again, 
the direction for which the fluctuation is canceled is 
\begin{eqnarray}
-\frac{{\rm d} \beta}{{\rm d} m}
=\frac{\bar{Q}_1^{\rm (hot)} - \bar{Q}_1^{\rm (cold)}}
{6(\bar{P}^{\rm (hot)} - \bar{P}^{\rm (cold)})}
\label{eq:minifull}
\end{eqnarray}
This is the same direction as the phase transition line. 
Therefore, the fluctuation of the reweighting factor along the phase 
transition line remains small, i.e. the statistical error does not increase 
so much. 

We obtained the same result as for the pure gauge theory on an anisotropic 
lattice, 
and this argument seems to be quite general for models with a first order 
phase transition, including models with chemical potential. 
However, there is a difference. Under the change of $a_{\sigma}/a_{\tau}$, 
any physics does not change along the $T_c$ line, but physical quantities, 
in general, depend on the quark mass. 
Although the dependence on $m$ might be much 
smaller than the dependence on $T/T_c$, if the fluctuation of the reweighting 
factor is completely canceled, any $m$-dependence is not obtained. 
In this discussion, we ignored the local fluctuation around the peaks 
in the hot and cold phase respectively, but the local fluctuation 
may play an important role for the $m$-dependence. 
Also, the sign problem for nonzero baryon density is caused by complex 
phase fluctuations of the reweighting factor (see Sec.~\ref{sec:phase}.), 
that is by the local fluctuations.
Hence the local fluctuation may, in particular, be important at non-zero 
baryon density.

\section{Quark mass reweighting for 2 flavor QCD}
\label{sec:massdir}

As we have seen in the previous two sections, the multi-parameter 
reweighting seems to be efficient to trace out the phase transition 
line in a wide range of the parameter space. 
One of the most interesting applications is finding the (pseudo-) critical 
line $(\beta_c)$ in the $(m, \beta)$ plane for 2 flavor or 2+1 flavor QCD. 
The phase transition for 2 flavor QCD at finite quark mass is expected 
to be crossover, which is not related to any singularity 
in thermodynamic observables, and that for 3 flavor QCD is crossover 
for quark masses larger than a critical quark mass, 
and is of first order for light quarks. 
The precise measurement of the (pseudo-) critical line is required for 
the extrapolation to the physical quark masses and for the study of 
universality class, e.g. to investigate for the 2 flavor case whether the 
chiral phase transition at finite temperature is in the same universality 
class as the 3-dimensional O(4) spin model or not.

In Ref.~\cite{us02}, we applied the reweighting method in the $(m, \beta)$ 
plane for 2 flavor QCD, and calculated the slope of the transition line, 
${\rm d} \beta_c /{\rm d} m$, where the reweighting factor with respect to 
quark mass was expanded into a power series and higher order terms which does
not affect the calculation of the slope were neglected. 
The results for ${\rm d} \beta_c /{\rm d} m$ compared well 
with the data of $\beta_c(m)$ obtained by direct calculations, 
without applying the reweighting method, 
demonstrating the reliability of a reweighting in 
a parameter of the fermion action. 
In this section, we discuss the relation between the fluctuation of 
the reweighting factor and the phase transition line in the $(m, \beta)$ 
plane for 2 flavor QCD at finite quark mass, i.e. at the crossover transition, 
by measuring the fluctuation in numerical simulations. 

For the presence of a direction for which the two reweighting factors 
from the gauge and the fermion action cancel, a correlation between 
these reweighting factors during Monte-Carlo steps is required. 
We estimate the correlation between these reweighting factors
using the configurations in Ref.~\cite{us02}. 
A combination of the Symanzik improved gauge action and 2 flavors of the p-4 improved 
staggered fermion action is employed \cite{HKS}. The parameters are 
$m_0=0.1, \beta_0=\{3.64, 3.645, 3.65, 3.655, 3.66, 3.665$ and $3.67\}$. 
The lattice size is $16^3 \times 4$. 7800-58000 trajectories are used 
for measurements at each $\beta$. 
The details are given in Ref.~\cite{us02}.\footnote{
The coefficient $c_3^F$ of the knight's move hopping term 
was incorrectly reported to be 1/96 in Ref.~\cite{us02}; 
its correct value is 1/48.} 

In the vicinity of the simulation point, the correlation of the 
reweighting factors can be approximated by
\begin{eqnarray}
\langle e^{F} e^{G} \rangle - \langle e^{F} \rangle \langle e^{G} \rangle 
\equiv \langle \Delta e^{F} \Delta e^{G} \rangle 
\approx \langle \Delta Q_1 \Delta P \rangle (m-m_0)(\beta-\beta_0)
 + \cdots , 
\end{eqnarray}
where $P=-\partial S_g/ \partial \beta$, 
$F=\sum_{n=1}^{\infty} Q_n (m-m_0)^n,$ and we denote 
$\Delta X \equiv X - \langle X \rangle$ for $X=\{P, Q_n,\cdots \}$. 
The $Q_n$ are obtained by 
\begin{eqnarray}
Q_1 = (N_{\rm f}/4) {\rm tr} M^{-1}, \ \
Q_2 = -(N_{\rm f}/8) {\rm tr} (M^{-1}M^{-1}), \cdots, 
\end{eqnarray}
for standard staggered fermions and also for p4-improved 
staggered fermions.
We calculate the value of 
$\langle \Delta Q_1 \Delta P \rangle \equiv 
\langle Q_1 P \rangle - \langle Q_1 \rangle \langle P \rangle$. 
The random noise method is used for the calculation of $Q_n$. 
The results for $\langle \Delta Q_1 \Delta P \rangle$
are listed in Table~\ref{tab:mass}.
We find strong correlation between the gauge and fermion parts of 
the reweighting factor. 

\begin{figure}[t]
\centerline{
\epsfxsize=9.5cm\epsfbox{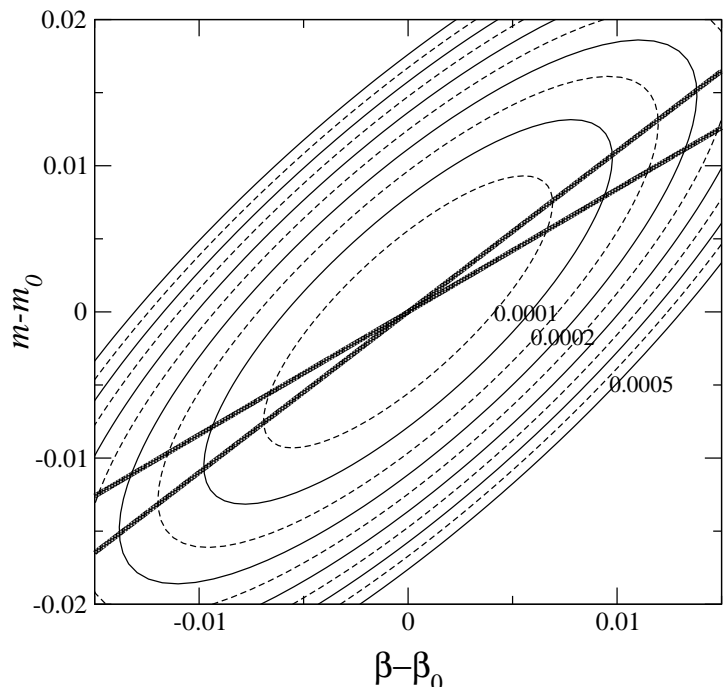}
}
\caption{
Contour plot of the standard deviation of the reweighting factor 
in the $(\beta, m)$ plane around $\beta_c$, 
$\beta_0=3.6492$, and $m_0=0.1$. 
Values in this figure are the square of the standard deviation 
divided by $N_{\rm site}$.
Bold lines show the upper bound and lower bound of 
$\partial \beta_c / \partial m$.
}
\vspace*{-4mm}
\label{fig:conm}
\end{figure}

Then we compute the total fluctuation of the reweighting factor as a function 
of $m$ and $\beta$ near the simulation point. 
Up to second order in $\beta - \beta_0$ and $m-m_0$, the square of 
the standard deviation is 
written as
\begin{eqnarray}
\langle [\Delta (e^{F} e^{G})]^2 \rangle 
& \approx &
\langle (\Delta Q_1)^2 \rangle (m-m_0)^2 
+ 2 \langle \Delta Q_1 \Delta P \rangle (m-m_0)(\beta-\beta_0) 
\nonumber \\ &&
+ \langle (\Delta P)^2 \rangle (\beta-\beta_0)^2 + \cdots . 
\end{eqnarray}
If we approximate in this form, lines of constant fluctuation 
(standard deviation) in the $(m, \beta)$ plane form ellipses. 
We also compute $\langle (\Delta Q_1)^2 \rangle $ and 
$\langle (\Delta P)^2 \rangle $, 
which are written in Table~\ref{tab:mass}. 
The values at $\beta_c=3.6492(22)$ 
are interpolated by applying the reweighting method for $\beta$ direction 
combining the data at seven simulation points \cite{Swen88}. 
The lines of constant fluctuation are drown in Fig.~\ref{fig:conm}. 
Numbers in this figure are the squares of the standard deviation 
divided by $N_{\rm site}$. 
It is found that these ellipses spread over one direction and 
the increase of the fluctuation is small along the direction. 
We also show the slope of the phase transition line by two lines: 
upper bound and lower bound of the derivative of $\beta_c$ with 
respect to $m$ obtained by measuring the peak position of the 
chiral susceptibility, 
${\rm d} \beta_c /{\rm d} m = 1.05(14)$ for $m_0=0.1$ \cite{us02}. 
We see that the directions of small fluctuations and of the phase 
transition are roughly the same. 
Since the fluctuation enlarges the statistical error of an observable, 
this figure can be also regarded as a map indicating the increase of 
the statistical error due to the reweighting. 
Therefore, we understand that the reweighting method can be applied 
in a wide range of parameters along the phase transition line if one 
performs simulations at the phase transition point. 

Moreover, it might be also important that these two directions are not exactly 
the same because, if the fluctuation is completely canceled along 
the transition line, no quantity can change, 
but the system should change as a function of quark mass even on 
the transition line, e.g. 
the chiral susceptibility should become larger as $m$ decreases.

\section{Chemical potential reweighting for 2-flavor QCD}
\label{sec:mudir}

\subsection{Correlation among the reweighting factors}
\label{sec:murewco}

Next, let us discuss the reweighting method for non-vanishing chemical 
potential. The reweighting method is really important for the study of 
finite density QCD since direct simulations are not possible for 
non-zero baryon density at present.
However, the complex measure problem (sign problem) is known to be 
a difficult problem. 
The reweighting factor for non-zero $\mu$ is complex. 
If the complex phase fluctuates rapidly and the reweighting factor 
changes sign frequently, the expectation values in Eq.(\ref{eq:rew})
become smaller than the error. Then the reweighting method breaks down. 
Therefore it is important to investigate the reweighting factor, 
including the complex phase, in practical simulations. 

First of all, we separate the fermion reweighting factor $e^{F}$
into an amplitude $|e^{F}|$ and a phase factor $e^{i \theta}$, 
and investigate the correlation among $|e^{F}|$, $e^{i \theta}$ and 
the gauge part $e^{G}$, where $e^{G}$ is real.
As is shown in Ref.~\cite{us02}, the phase factor and the amplitude 
can be written as the odd and even terms of the Taylor 
expansion of $\ln \det M$, respectively, since the odd terms are 
purely imaginary and the even terms are real at $\mu=0$. 
Denoting $F=\sum_{n=1}^{\infty} R_n \mu^n,$ 
\begin{eqnarray}
|e^{F}|=\exp\{\sum_{n=1}^{\infty} {\rm Re} 
(R_{2n}) \mu^{2n}\} \ \ {\rm and} \ \
e^{i \theta}=\exp\{i \sum_{n=1}^{\infty} {\rm Im} 
(R_{2n-1}) \mu^{2n-1}\}.
\end{eqnarray}
We study the correlation among these factors in the vicinity of 
the simulation point $(\beta_0, \mu_0=0)$. 
Up to $O(\beta-\beta_0, \mu^2)$, the reweighting factor is 
\begin{eqnarray}
e^{i \theta} |e^{F}| e^{G} \approx 1 + R_1 \mu 
+ (R_1^2/2) \mu^2 + R_2 \mu^2 + P (\beta-\beta_0) +\cdots . 
\end{eqnarray}
We compute the correlations, 
$\langle \Delta (R_1^2/2) \Delta P \rangle,$
$\langle \Delta R_2 \Delta P \rangle,$ and 
$\langle \Delta (R_1^2/2) \Delta R_2 \rangle$ at $\mu=0$, 
which correspond to the correlations of $(e^{i \theta}, e^{G})$, 
$(|e^{F}|, e^{G})$ and $(e^{i \theta}, |e^{F}|)$, 
respectively. 
$\Delta X \equiv X - \langle X \rangle$ for $X=\{P, R_n,\cdots \}$. 
Here, $\langle \Delta R_1 \Delta P \rangle$ and
$\langle \Delta R_1 \Delta R_2 \rangle$ are zero 
at $\mu=0$ because $R_1$ is purely imaginary.
The $R_n$ are obtained by 
\begin{eqnarray}
R_1 &=& \frac{N_{\rm f}}{4}
\frac{\partial \ln \det M}{\partial \mu} 
= \frac{N_{\rm f}}{4}
{\rm tr} \left( M^{-1} \frac{\partial M}{\partial \mu} \right), 
\label{eq:dermu1} \\
R_2 &=& \frac{N_{\rm f}}{4} \frac{1}{2}
\frac{\partial^2 \ln \det M}{\partial \mu^2} 
= \frac{N_{\rm f}}{4} \frac{1}{2} \left[ 
{\rm tr} \left( M^{-1} \frac{\partial^2 M}{\partial \mu^2} \right)
 - {\rm tr} \left( M^{-1} \frac{\partial M}{\partial \mu}
                   M^{-1} \frac{\partial M}{\partial \mu} \right) \right], 
\label{eq:dermu2} \\
R_3 &=& \frac{N_{\rm f}}{4} \frac{1}{3!}
\frac{\partial^3 (\ln \det M)}{\partial \mu^3} 
= \frac{N_{\rm f}}{4} \frac{1}{3!} \left[
{\rm tr} \left( M^{-1} \frac{\partial^3 M}{\partial \mu^3} \right)
 -3 {\rm tr} \left( M^{-1} \frac{\partial M}{\partial \mu}
 M^{-1} \frac{\partial^2 M}{\partial \mu^2} \right) \right. \nonumber \\
&& \ \ \left. +2 {\rm tr} \left( M^{-1} \frac{\partial M}{\partial \mu}
        M^{-1} \frac{\partial M}{\partial \mu}
        M^{-1} \frac{\partial M}{\partial \mu} \right) \right], 
\label{eq:dermu3}
\end{eqnarray}
for staggered type fermions. Details of the calculation are 
given in Ref.~\cite{us02}.

We use the configurations in Ref.~\cite{us02}, again, generated 
by the $N_{\rm f}=2$ $p4$-improved action on a $16^3 \times 4$ lattice. 
We generated $20000$-$40000$ trajectories for $m_0=0.1$, 
$\beta_0=\{3.64, 3.65, 3.66$ and $3.67\}$. 
The results are summarized in Table~\ref{tab:mu}. 
We find that the correlation between $|e^{F}|$ and 
$e^{G}$ is very strong in comparison with the other 
correlations, which means that the contribution to an observable can be 
separated into two independent parts: from $e^{i \theta}$, and 
from a combination of $|e^{F}| \times e^{G}.$

To make the meaning of this result clearer, we consider the following 
partition function, introducing two different $\mu$; $\mu_o$ and $\mu_e$, 
\begin{eqnarray}
\label{eq:par}
{\cal Z}= \int D U e^{R_1 \mu_o +R_3 \mu_o^3 +\cdots} 
e^{R_2 \mu_e^2 +R_4 \mu_e^4 +\cdots} 
(\det M|_{\mu=0})^{N_{\rm f}/4} e^{-S_g} 
\label{eq:modact} 
\end{eqnarray}
Then, at $\mu=0$, 
\begin{eqnarray}
\langle \Delta R_1^2 \Delta P \rangle 
= \langle (\Delta R_1)^2 \Delta P \rangle 
&=& \frac{\partial^3 \ln {\cal Z}}{\partial \mu_o^2 \partial \beta} 
= N_{\rm site} \frac{\partial (\chi_q a^2 - 4\chi_{\rm IV} a^2)}
{\partial \beta}, \\
2 \langle \Delta R_2 \Delta P \rangle 
&=& \frac{\partial^3 \ln {\cal Z}}{\partial \mu_e^2 \partial \beta} 
= N_{\rm site} \frac{\partial (4\chi_{\rm IV} a^2)}{\partial \beta}, 
\end{eqnarray}
where $\chi_q$ and $\chi_{\rm IV}$ are the quark number 
susceptibility and iso-vector quark number susceptibility \cite{Gottlieb}: 
\begin{eqnarray}
\frac{\chi_q}{T^2}
&=& \left( \frac{\partial}{\partial (\mu_{\rm u}/T)}
+\frac{\partial}{\partial (\mu_{\rm d}/T)} \right)
\frac{n_{\rm u} + n_{\rm d}}{T^3}, \label{eq:chisq}\\
\frac{4\chi_{\rm IV}}{T^2}
&=& \left( \frac{\partial}{\partial (\mu_{\rm u}/T)}
-\frac{\partial}{\partial (\mu_{\rm d}/T)} \right)
\frac{n_{\rm u} - n_{\rm d}}{T^3}. \label{eq:chist}
\end{eqnarray}
We choose the same chemical potential for up and down quarks: 
$\mu_{\rm u}=\mu_{\rm d} \equiv \mu_q$.
$n_{\rm u(d)}$ is the number density for up (down) quark: 
$n_{\rm u(d)}/T^3=\partial(p/T^4)/\partial(\mu_{\rm u(d)}/T)$. 
Quark and baryon number susceptibilities are related 
by $\chi_{\rm B}\equiv\partial n_{\rm B}/\partial\mu_{\rm B}=3^{-2}\chi_q$.
If we impose a chemical potential with opposite sign for up and 
down quarks: $\mu_{\rm u}=-\mu_{\rm d} \equiv \mu_{\rm IV}/2,$ 
called ``iso-vector chemical potential", the Monte-Carlo method is 
applicable since the measure is not complex \cite{SS,KogSin}. 
For this model, the iso-vector quark number susceptibility $\chi_{\rm IV}$ 
in Eq.(\ref{eq:chist}) is the quark number susceptibility instead of 
Eq.(\ref{eq:chisq}). 

The result in Table~\ref{tab:mu} means that 
\begin{eqnarray}
\frac{\partial^3 \ln {\cal Z}}{\partial \mu_o^2 \partial \beta} 
 \ll \frac{\partial^3 \ln {\cal Z}}{\partial \mu_e^2 \partial \beta},
\end{eqnarray}
i.e. $\mu$ in the phase factor $(\mu_o)$ does not contribute to the 
$\beta$-dependence of ${\cal Z}$ near $\mu=0$, hence $\mu$ in the amplitude 
$(\mu_e)$ is more important for the determination of $\beta_c$ by measuring 
the $\beta$-dependence of thermodynamic quantities.
Moreover, these correlations have a relation with the slope of 
$\chi_q$ and $4\chi_{\rm IV}$ in terms of $\beta$. 
Since $\chi_q - 4\chi_{\rm IV}$ is known to be small at $\mu=0$ 
\cite{Gottlieb},\footnote{
However, as $\mu$ increases, the difference between 
$\chi_q$ and $4\chi_{\rm IV}$ becomes sizeable \cite{us03}, which might 
be related to only $\chi_q$ being expected to have 
a singularity at the critical endpoint \cite{HS}.} 
this result may not change even for small quark mass.
Also, the result in Ref.~\cite{SNT} suggests that the effect of 
the phase factor, 
i.e. of $\mu_o$, on physical quantities is small. 

\paragraph*{Iso-vector chemical potential}
Furthermore, we discuss the model with iso-vector chemical potential. 
In Ref.~\cite{us02}, we discussed the difference in the curvature 
of the phase transition to that of the usual chemical potential.
Because we expect that at $T=0$ pion condensation happens around 
$\mu_{\rm q} \approx m_{\pi}/2$, and that the phase transition line 
runs to that point directly, the curvature of the transition line for 
iso-vector $\mu$ should be much larger than that for usual $\mu$, 
since $m_{\pi}/2 \ll m_{\rm N}/3$. 
However, as we discussed above, $\mu_o$ in Eq.(\ref{eq:par}) does not 
contribute to the shift of $\beta_c$ near $\mu=0$ and the difference from 
the usual $\mu$ is only in $\mu_o$, i.e. $\mu_o=0$ for the iso-vector case. 
Therefore the difference in the curvature might be small and 
the naive picture seems to be wrong.
In practice, our result at small $\mu$ using the method in Ref.~\cite{us02} 
supports that. 
Moreover, Kogut and Sinclair \cite{Kog02} showed that $\beta_c$ from 
chiral condensate measurements is fairly insensitive to $\mu$ for small 
$\mu$ by direct simulations with iso-vector $\mu$.

\subsection{Fluctuation of the reweighting factor}
\label{sec:murewfl}

Next, we estimate the fluctuation of the reweighting factor. 
As we have seen above, the fluctuation of the reweighting factor is 
separated into the complex phase factor of $e^{\Delta F}$ and 
the other part, and these are almost independent. 
Moreover, this implies the absolute value of $e^{\Delta F}$ 
is important for the determination of $\beta_c$. 
The amplitude of the fermionic part $|e^{\Delta F}|$ and 
the gauge part $e^{\Delta G}$ are strongly correlated, and then 
the variation of the total fluctuation of these parts 
in the parameter space is not simple. Because the total fluctuation 
is related to the applicability range of the reweighiting method, 
here, we compute the standard deviation of $|e^{F}| e^{G}$ 
to estimate the fluctuation, and also discuss the relation to the 
phase transition line. 
The complex phase fluctuation $e^{i \theta}$ will be discussed 
in the next section separately. 

\begin{figure}[t]
\centerline{
\epsfxsize=9.5cm\epsfbox{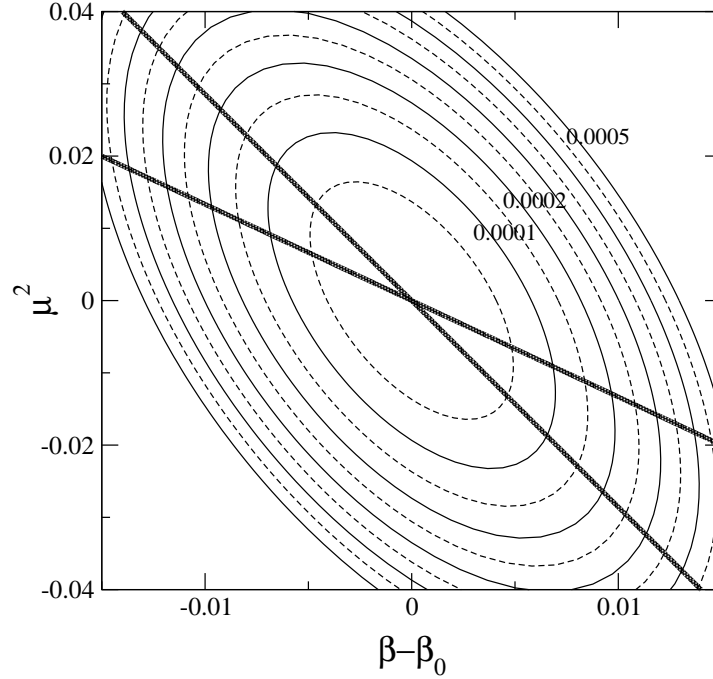}
}
\caption{
Contour plot of the standard deviation of the reweighting factor 
in the $(\beta, \mu^2)$ plane around $\beta_c$, 
$\beta_0=3.6497$, and $m_0=0.1$. 
Bold lines show upper bound and lower bound of 
$\partial \beta_c / \partial (\mu^2)$.
}
\vspace*{-4mm}
\label{fig:conmu}
\end{figure}

Up to the leading order of $\beta-\beta_0$ and $\mu^2$, 
the square of the standard deviation is obtained by 
\begin{eqnarray}
\langle [\Delta (|e^{F}| e^{G}) ]^2 \rangle \approx 
\langle (\Delta R_2)^2 \rangle \mu^4 
+2 \langle \Delta R_2 \Delta P \rangle \mu^2 (\beta-\beta_0) 
+\langle (\Delta P)^2 \rangle (\beta-\beta_0)^2 + \cdots. 
\end{eqnarray}
Then, the line of constant fluctuation is an ellipse in this 
approximation. We show the contour lines in Fig.~\ref{fig:conmu}. 
The susceptibilities and the correlation of $R_2$ and $P$, 
$\langle (\Delta R_2)^2 \rangle$, $\langle (\Delta P)^2 \rangle$ 
and $\langle \Delta R_2 \Delta P \rangle$, 
are summarized in Table~\ref{tab:mu}. 
The values at the phase transition point, 
$\beta_c=3.6497(16)$, are computed by the reweighting method 
for the $\beta$ direction using the data at four $\beta$ points. 
Numbers in this figure are the squares of the standard deviation 
divided by $N_{\rm site}$. 
We also denote the lower and upper bounds of 
$\partial^2 \beta_c / \partial \mu^2=-1.1(4)$ by bold lines, 
which are obtained by the measurement of the chiral susceptibility 
\cite{us02}.
We find that there exists a direction along which the increase of 
the fluctuation is relatively small, and 
this direction is roughly parallel to the phase transition line.
Because we expect that physics is similar along the transition line, 
if we consider that $|e^{F}| e^{G}$ is the important part 
for the calculation of $\beta_c$, this result is quite reasonable. 

As well as in the $(m, \beta)$ plane,
the fluctuation of the reweighting factor is small along the phase 
transition line in the $(\mu, \beta)$ plane, and the reweighting 
method seems to be efficient to trace out the phase transition line. 
This must be a reason why the phase transition line can be determined for 
rather large $\mu$ in Ref.~\cite{FK}. 
However, in this discussion, we omitted the complex phase fluctuation, 
and the phase fluctuation is the most important factor for the sign problem. 
As we will discuss in the next section, the value of $\mu$ for which  
the sign problem arises depends strongly on the lattice size.
The sign problem is not very severe for small lattices such as $4^4$, 
$6^3 \times 4$ and $8^3 \times 4$ lattices employed in Ref.~\cite{FK}, 
which is also an important reason for their successful calculation. 

\begin{figure}[t]
\centerline{
\epsfxsize=10.0cm\epsfbox{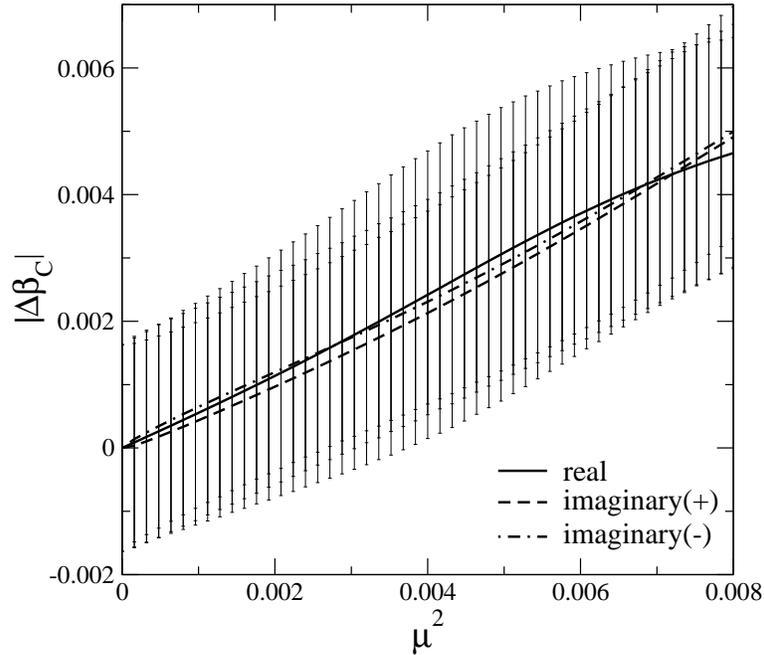}
}
\caption{
$|\beta_c(\mu)-\beta_c(0)| \equiv |\Delta \beta_c|$ 
for real $\mu$ and imaginary $\mu$. 
Solid line is the result for real $\mu$. 
Dashed and dot-dashed lines are the results of $\mu \to i\mu$ and 
$\mu \to -i\mu$ respectively. 
}
\vspace*{-4mm}
\label{fig:ivmu}
\end{figure}

\paragraph*{Imaginary chemical potential}
In Fig.~\ref{fig:conmu}, we show also the region for $\mu^2<0$, 
i.e. imaginary $\mu$. 
de Forcrand and Philipsen \cite{dFP} computed 
$\partial^2 \beta_c / \partial \mu^2$ performing simulations with 
imaginary $\mu$, assuming that $\beta_c$ is an even function in 
$\mu$ and analyticity in that region. 
(Also in Ref.~\cite{dEL} for $N_{\rm f}=4$.)
The $\beta_c (\mu)$ for imaginary $\mu$ shifts the opposite 
direction to that for real $\mu$ as $\mu$ increases, 
but the absolute value of the second derivative, 
$\partial^2 \beta_c / \partial \mu^2$, is expected to be the same.
Here, we confirm whether the results of 
$|\partial^2 \beta_c / \partial \mu^2|$ obtained by real and imaginary 
$\mu$ are consistent or not by the method in Ref.~\cite{us02}.  
We replace $\mu$ by $i\mu$ or $-i\mu$ and reanalyze for imaginary $\mu$. 
In Ref.~\cite{us02}, the reweighting factor has been obtained 
in the form of the Taylor expansion in $\mu$ up to $O(\mu^2)$, 
and the replacement is easy. 
We determined $\beta_c$ by the peak position of the chiral susceptibility, 
using the data at $m_0=0.1$ in Ref.~\cite{us02}. 
The results of $|\beta_c(\mu) -\beta_c(0)|$ are shown 
in Fig.~\ref{fig:ivmu}. Errors by the truncation of the Taylor expansion 
terms are $O(\mu^4)$.
The solid line is the result for real $\mu$. 
The results of $\mu \to i\mu$ and $\mu \to -i\mu$ are dashed and 
dot-dashed lines respectively. 
The slope at $\mu=0$ is $-(\partial^2 \beta_c / \partial \mu^2)/2$. 
We find that these results of the slope for real and imaginary $\mu$ 
are consistent.
It has been also discussed for measurements of spatial correlation lengths 
to confirm the reliability of the analytic continuation 
from imaginary chemical potential \cite{Laine}.

\section{Complex phase fluctuation}
\label{sec:phase}

Finally, it is worth to discuss the complex phase fluctuation in order to 
know the region of applicability of generic reweighting approaches.
If the reweighting factor in Eq.(\ref{eq:rew}) changes 
sign frequently due to the complex phase of the quark determinant, 
then both numerator and denominator of Eq.(\ref{eq:rew}) become
very small in comparison with the statistical error. 
Of course, the complex phase starts from zero at $\mu=0$ but grows 
as $\mu$ increases. It is important to establish at 
which value of $\mu$ the sign problem becomes severe.

As discussed in the previous section, the phase can be expressed 
using the odd terms of the Taylor expansion of $\ln \det M$. 
The complex phase is 
\begin{eqnarray}
\label{eq:phase}
\theta = {\rm Im} (R_1 \mu + R_3 \mu^3 + R_5 \mu^5 + \cdots ).
\end{eqnarray}
The explicit expression for $R_1$ and $R_3$ are given in 
Eqs.~(\ref{eq:dermu1}) and (\ref{eq:dermu3}). 

Because the sign of the real part of the complex phase changes at 
$\theta=\pi/2$, 
the sign problem occurs when the typical magnitude of $\theta$ 
becomes larger than $\pi/2$. 
We use the point at which the magnitude 
of the phase reaches the value $\pi/2$ as a simple criterion to 
estimate the parameter range in which reweighting methods will start 
to face serious sign problems. 
If the sign problem arises at small $\mu$, which is expected to happen 
for a large lattice, the first term in Eq.(\ref{eq:phase}) is most 
important. Then we can estimate the applicability range by evaluating 
the fluctuation of $R_{1}$. 
Moreover we expect naively that the magnitude of 
${\rm tr}[M^{-1} (\partial M/ \partial \mu) \cdots]$ 
is proportional to $N_{\rm site}$, therefore the value of $\mu$ 
at which the sign problem arises decreases roughly in inverse 
proportion to the number of site $N_{\rm site}$. 
Also, the situation is different on lattices of moderate size. 
In Ref.~\cite{dFKT}, it is shown that the first term 
in Eq.(\ref{eq:phase}) is dominant for $\mu=0.1$ and $0.2$ 
but the higher order terms cannot be neglected for $\mu \simgt 0.3$, 
by calculating the complex phase without the approximation 
by the Taylor expansion. If the higher order 
terms are not negligible, the volume dependence is not simple. 
E.g. in the case that the term of $O(\mu^3)$ plays an important role 
for the determination of the applicability range of the reweighting method, 
the applicability range is expected to decrease in proportion to 
$N_{\rm site}^{-1/3}$, and similarly $N_{\rm site}^{-1/5}$ 
for the case that the $O(\mu^5)$ term is important. 

We consider the leading term and the next leading term of the complex phase. 
The expectation value of $\theta$ must be zero at $\mu=0$ because the 
partition function is real. Although the average of the phase is zero, 
its fluctuations remain important. 
We investigate the standard deviation of $\theta$ up to $O(\mu^3)$, 
${\rm STD}(\theta) \equiv \sqrt{\left\langle \theta^2 \right\rangle - 
\left\langle \theta \right\rangle^2}$, 
using configurations generated on a $16^3 \times 4$ lattice 
for the study of Ref.~\cite{us03}, 
and also the standard deviations of ${\rm Im} (R_1)$ and 
${\rm Im} (R_3)$ are computed.

The random noise method is used to calculate $\theta$ for each 
configuration. Then, the value of $\theta$ contains an error due to 
the finite number of noise vectors $N_{\rm n}$. 
To reduce this error, we treat the calculation of 
$\langle \theta^2 \rangle$ more carefully. 
Since the noise sets for the calculation of the two $\theta$ in the 
product must be independent, we subtract the contributions 
from using the same noise vector for each factor. 
Details are given in the Appendix of Ref.~\cite{us02}. 
By using this method, we can make the $N_{\rm n}$-dependence of 
$\langle \theta^2 \rangle$ much smaller than that by the naive 
calculation from rather small $N_{\rm n}$, 
hence it may be closer to the $N_{\rm n}=\infty$ limit. 
We took $N_{\rm n}=50$ for this calculation. 

\begin{figure}[t]
\centerline{
\epsfxsize=9.5cm\epsfbox{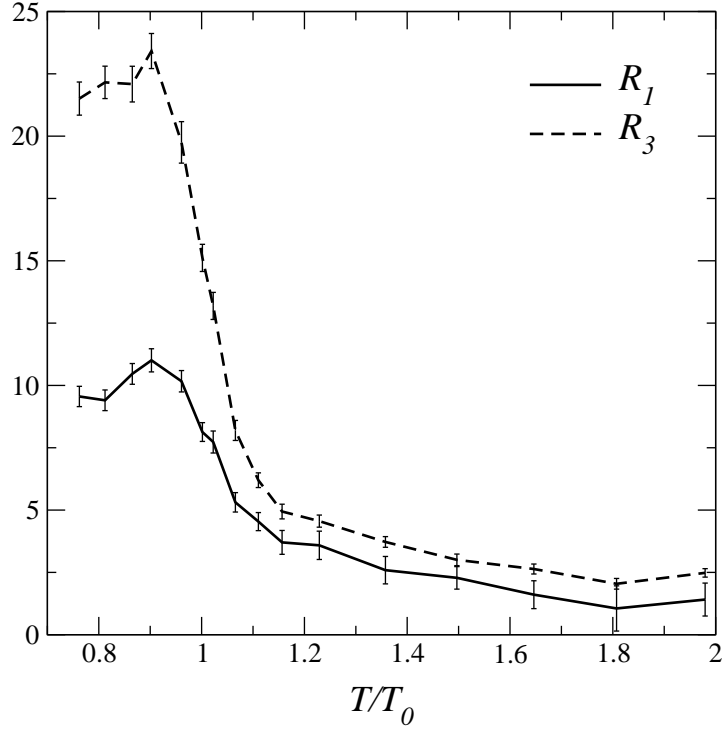}
}
\caption{
Standard deviation of ${\rm Im}(R_1)$ and ${\rm Im}(R_3)$.
$T_0$ is $T_c$ at $\mu=0$.
}
\vspace*{-4mm}
\label{fig:stdr13}
\end{figure}

In Fig.~\ref{fig:stdr13}, we plot the standard deviations of the $O(\mu)$ 
term, ${\rm Im} (R_1)$, and the $O(\mu^3)$ term, ${\rm Im} (R_3)$, 
for $N_{\rm f}=2$, $m=0.1$. 
The horizontal axis is temperature normalized by $T_c$ at $\mu=0$ $(T_0)$. 
The temperature scale is determined from the string tension data in 
Ref.~\cite{KLP} with the fit ansatz of Ref.~\cite{Chris}.
The fluctuations of these terms are almost of the same magnitude and 
both of them are small in the high temperature phase, 
hence the sign problem is not serious in the high temperature phase. 
We, moreover, confirm that the $O(\mu)$ term is dominant 
around $\mu \equiv \mu_{\rm q} a \sim 0.1$, as suggested by Ref.~\cite{dFKT}, 
and the approximation by $O(\mu)$ term for the discussion of the applicability 
range in Ref.~\cite{us02} is valid for the $16^3 \times 4$ lattice. 
This suggests the applicability range decreases roughly in proportion 
to $N_{\rm site}^{-1}$. However, in general, 
the magnitude of the fluctuation (standard deviation) of 
$R_1/N_{\rm site}$  
changes as a function of the volume, hence the detailed finite size 
analysis is necessary to investigate the volume dependence 
of the applicability range more precisely. 

Recently, analysis of the volume dependence of the applicability range 
has been reported in Ref.~\cite{Fodor04}. 
Their numerical result of the applicability range is 
in proportion to $N_{\rm site}^{-1/3}$. 
This is much better than $N_{\rm site}^{-1}$.
Because their estimations are based on simulations on $6^3 \times 4$, 
$8^3 \times 4$, $10^3 \times 4$ and $12^3 \times 4$ lattices, 
the applicability ranges are relatively large, 
and the higher order terms in $\mu$ should be important for large $\mu$. 
Therefore the result of the volume dependence, 
$\mu \sim N_{\rm site}^{-1/3}$, is reasonable for their lattice size, 
but may will change on large lattices. 

\begin{figure}[t]
\centerline{
\epsfxsize=10.0cm\epsfbox{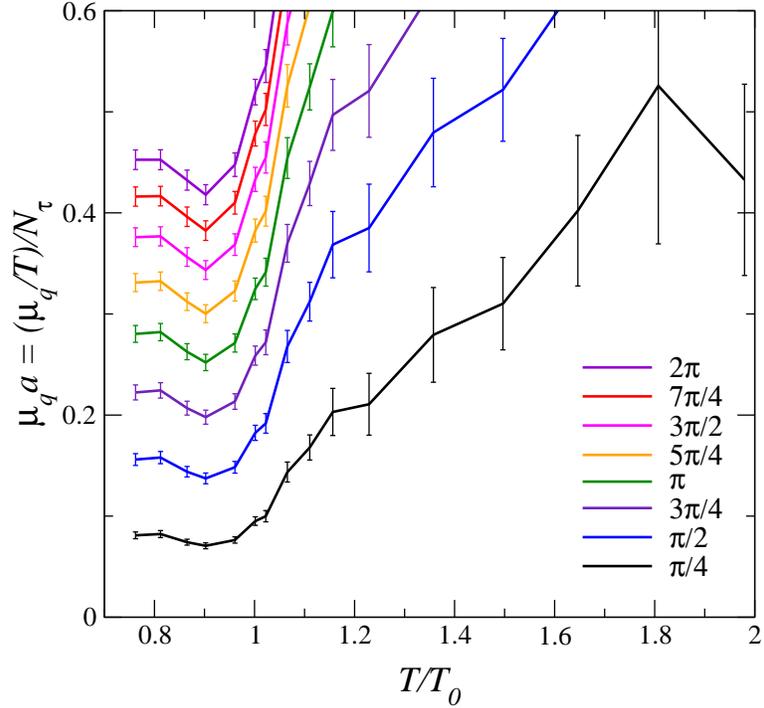}
}
\caption{
Contour plot of the complex phase fluctuation ${\rm STD}(\theta)$ 
in the $(T/T_0, \mu_q a \equiv (\mu_q/T)N_{\tau}^{-1})$ plane. 
The complex phase $\theta$ contains $O(\mu^5)$ error. 
$T_0$ is $T_c$ at $\mu=0$. $N_{\tau}=4$.
}
\vspace*{-4mm}
\label{fig:contpf}
\end{figure}

We also show the contour plot for 
${\rm STD}(\theta)=
\{\pi/4, \pi/2, 3\pi/4, \pi, 5\pi/4, 3\pi/2, 7\pi/4, 2\pi \}$ 
in Fig.~\ref{fig:contpf}. 
The error of the contour is estimated by the jack knife method. 
At the interesting regime for the heavy-ion collisions: 
$\mu_{\rm q}/T_c \approx 0.1 $ for RHIC and 
$\mu_{\rm q}/T_c \approx 0.5$ for SPS, 
the fluctuation is smaller than $\pi/2$ in the whole range of $T$. 
Therefore, the reweighting method seems to be applicable for the quantitative 
study for the heavy-ion collisions, which is an encouraging result. 
Also, we find that a point around $T/T_0=0.9$ looks singular. 
Because we expect the fluctuation of the system diverges at a critical point, 
this might be related to the presence of a critical endpoint. 
The large fluctuation around $T/T_0=0.9, \mu_q a>0.5$ 
is occurred by the $O(\mu^3)$ term being large around there. 
This might be corresponding that, in general, the contribution from the 
higher order terms of the Taylor expansion becomes larger, 
as the critical point is approached, 
and the expansion series does not converge near the critical point. 
Plus sign and minus sign appear with almost equal probability, i.e. 
$\langle e^{i \theta} \rangle$ is almost zero, when the 
standard deviation of $\theta$ is larger than $\pi$. 
The value of $\mu_q/T=\mu N_{\tau}$ at which 
the standard deviation of $\theta$ is $\pi$ around $T_c$ is 
of the order $\mu_q/T \sim O(1)$. 
However, we should notice that the complex phase, again, is very 
sensitive to the lattice size $N_{\rm site}$. 
For small lattice size, the sign problem is not severe and 
the reweighting method can be used for considerably large $\mu$, 
however the applicability range of the reweighting method will 
be narrower for a lattice with a size larger than $16^3 \times 4$. 
Also the analysis of quark mass dependence must be important as a part 
of future investigations.

\section{Conclusions}
\label{sec:concl}

At present, the reweighting method is an important approach to the study of 
QCD at finite baryon density. 
We discussed the applicability range of the reweighting method with 
multi-parameters. 
The fluctuation of the reweighting factor during Monte-Carlo steps 
is a cause of the increase of the statistical error due to the reweighting, 
and the magnitude of the fluctuation determines the applicable range. 

For a simulation of SU(3) pure gauge theory 
at the first order phase transition point on an anisotropic lattice, 
the fluctuation is minimized along the phase transition line 
in the parameter space of an anisotropic lattice, 
if we assume the pressure in hot and cold phases to be balanced.
This argument can also be applied in full QCD, if the theory has a first order 
phase transition, and the same relation between the fluctuation of the 
reweighting factor and the phase transition line is obtained. 
This suggests that the multi-parameter reweighting method is an efficient 
method for trace out the phase transition line in multi-parameter space. 
Moreover, the measurement of thermodynamic quantities on 
the phase transition line is important for the finite size scaling analysis 
to discuss the universality class. 
The reweighting method is useful for this purpose.

We also measured the fluctuation of the reweighting factor 
in numerical simulations of 2 flavor QCD for the cases of the reweighting 
in quark mass and chemical potential. 
There exists a direction of small fluctuation in the 
$(m, \beta)$ plane and it is roughly the same direction as 
that of the phase transition line. 
The fluctuation of the reweighting factor with respect to chemical 
potential can be separated into two parts: the complex phase 
factor of the fermion part and, on the other hands, the absolute value of 
the fermion part and the gauge part. 
These two parts fluctuate almost independently 
during Monte-Carlo steps. This implies that the phase factor of 
the fermion part does not influence the shift of $\beta_c$ with increasing 
$\mu$ at small $\mu$, and also explains why the difference between the phase 
transition lines for usual chemical potential and iso-vector chemical 
potential is small at low density. 
If we neglect the complex phase factor, the increase of the fluctuation 
is also small along the phase transition line in the $(\mu^2, \beta)$ plane, 
as well as in the $(m, \beta)$ plane. 

The value of $\mu$ for which the sign problem arises decreases as 
the lattice size $N_{\rm site}$ increases, 
hence simulations at $\mu \neq 0$ are more difficult for 
larger lattices, even if the fluctuation of the absolute value of the 
reweighting factor is small along the phase transition line. 
The complex phase fluctuation is measured on a $16^3 \times 4$ lattice. 
The sign problem is not serious in the high temperature phase, but 
around the phase transition point it becomes serious gradually from 
this lattice size. For small lattices, 
the sign problem is not severe for the study at low density, 
and also for the $16^3 \times 4$ lattice, the applicability range of the 
reweighting method covers the interesting regime for heavy-ion collisions. 
Also, the behavior of the complex phase fluctuation around the transition 
point suggests a critical endpoint in the region of $\mu_q/T_c \sim O(1)$.

\section*{Acknowledgments}
I wish to thank C.R.~Allton, S.J.~Hands, O.~Kaczmarek,
F.~Karsch, E.~Laermann, and C.~Schmidt 
for fruitful discussion, helpful comments and 
allowing me to use the data in Refs.~\cite{us02} and \cite{us03}. 
I also thank Y.~Iwasaki, K.~Kanaya and T.~Yoshi\'{e} for giving me 
the data in Ref.~\cite{QCDPAX} and Ph.~de~Forcrand for useful comments
on this manuscript.
This work is supported by BMBF under grant No.06BI102 
and PPARC grant PPA/G/S/1999/00026.

\begin{table}[thb]
\caption{Correlation and susceptibilities of $Q_1$ and $P$. 
$N_{\rm site} =16^3 \times 4.$
$\beta_c=3.6492(22).$}
\label{tab:mass}
\begin{center}
\begin{tabular}{cccc}
$\beta$ & 
$\langle \Delta Q_1 \Delta P \rangle N_{\rm site}^{-1}$ & 
$\langle (\Delta Q_1)^2 \rangle N_{\rm site}^{-1}$ & 
$\langle (\Delta P)^2 \rangle N_{\rm site}^{-1}$ \\
\hline
3.640 & -1.44(6)  & 1.29(6)  & 2.67(7)  \\
3.645 & -1.80(13) & 1.69(13) & 2.99(14) \\
3.650 & -1.70(6)  & 1.58(6)  & 2.89(7)  \\
3.655 & -1.76(14) & 1.65(13) & 2.92(15) \\
3.660 & -1.60(6)  & 1.49(6)  & 2.77(7)  \\
3.665 & -1.36(13) & 1.19(12) & 2.58(14) \\
3.670 & -1.52(7)  & 1.41(8)  & 2.68(8)  \\
\hline
3.6492 & -1.74(4) & 1.61(3) & 2.92(5) \\
\end{tabular}
\end{center}
\end{table}

\begin{table}[thb]
\caption{Correlations and susceptibilities 
among $R_1^2, R_2$ and $P$. 
$N_{\rm site} =16^3 \times 4.$
$\beta_c=3.6497(16).$}
\label{tab:mu}
\begin{center}
\begin{tabular}{cccccc}
$\beta$ & 
$\langle \Delta (R_1^2/2) \Delta P \rangle N_{\rm site}^{-1}$ & 
$\langle \Delta R_2 \Delta P \rangle N_{\rm site}^{-1}$ & 
$\langle \Delta (R_1^2/2) \Delta R_2 \rangle N_{\rm site}^{-1}$ &
$\langle (\Delta R_2)^2 \rangle N_{\rm site}^{-1}$ &
$\langle (\Delta P)^2 \rangle N_{\rm site}^{-1}$ \\
\hline
3.64 & 0.006(29) & 0.312(33) & 0.034(10) & 0.216(16) & 2.62(10) \\
3.65 & 0.059(21) & 0.434(29) & 0.056(10) & 0.254(14) & 2.87(8)  \\
3.66 & 0.055(15) & 0.410(26) & 0.022(5)  & 0.231(11) & 2.75(8)  \\
3.67 & 0.032(15) & 0.397(28) & 0.031(5)  & 0.219(13) & 2.68(8)  \\
\hline
3.6497 & 0.059(13) & 0.495(19) & 0.050(7) & 0.267(9) & 2.98(7) \\
\end{tabular}
\end{center}
\end{table}

\end{document}